\documentclass[article,twocolumn,floatfix]{revtex4}
\usepackage{times}

\usepackage{graphicx}
  \graphicspath{{./Eps/}}
\usepackage{amsmath}
\usepackage{amssymb}

\begin{document}

\title{States-conserving density of states for Altshuler-Aronov effect: Heuristic derivation}

\author{A. \surname{Mo\v{s}kov\'{a}}}
\affiliation{Institute of Electrical Engineering, Slovak Academy of Sciences, 841 04 Bratislava, Slovakia}
\author{M. \surname{Mo\v{s}ko}}
\email{martin.mosko@savba.sk}

\affiliation{Faculty of Mathematics, Physics and Informatics, Comenius University, 842 48  Bratislava, Slovakia}
\affiliation{Institute of Electrical Engineering, Slovak Academy of Sciences, 841 04 Bratislava, Slovakia}

\begin{abstract}
Altshuler and Aronov (AA) have shown that the electron-electron interaction in a weakly-disordered metal suppresses the single-particle density of states (DOS) in the vicinity of the Fermi level ($E_F$).
According to the AA theory the suppressed DOS exhibits the energy dependence $\propto \sqrt {|E-E_F|}$ valid for $|E-E_F|$ smaller than a certain correlation energy $U_{co}$.
  Recent experiments have shown that at energies larger than $U_{co}$ the DOS exhibits a states-conserving dependence on energy, namely, 
  the states removed from near the Fermi level are found at energies above $U_{co}$ in the energy range of about $3 U_{co}$. 
  In this work the AA effect is studied beyond the low energy limit theoretically. We consider the AA model
in which the electrons interact via the statically screened Coulomb interaction and the modification of the DOS is due 
to the exchange part of the electron self-energy. We derive the states-conserving DOS heuristically. 
Namely, we show that the self-energy consists of a diverging part (which we skip on physical grounds) and of the small part of the order of the pair Coulomb energy. 
This small part gives the states-conserving DOS which is in qualitative accord with experimental observations at energies above $U_{co}$ and which reproduces the AA result at energies below $U_{co}$.
\end{abstract}


\maketitle

\section{Introduction}
Altshuler and Aronov (AA) have shown  \cite{Altshuler1,Altshuler3,Altshuler4} that the electron-electron (e-e) interaction in a weakly-disordered three-dimensional ($3D$) metal
suppresses the single-particle density of states (DOS) in the vicinity of the Fermi level ($E_F$).
Specifically, the AA theory predicts for the suppressed DOS the energy dependence $\propto \sqrt {|E-E_F|}$ which is valid for $|E-E_F| \lesssim U_{co}$
 where $U_{co}$ is a characteristic correlation energy.
The DOS $\propto \sqrt{|E-E_F|}$ at energies $|E-E_F| \lesssim U_{co}$ was observed by tunneling spectroscopy 
\cite{Abeles,Dynes,McMillan2,ImryOvadyahu,Schmitz1,Schmitz2,Escudero,Teizer,Mazur,Luna2014,Luna2015} and by
photoemission spectroscopy \cite{Kobayashi}.
Some experiments \cite{Schmitz1,Schmitz2,Escudero,Mazur,Kobayashi} studied the DOS also for  $|E-E_F| > U_{co}$. In particular, the aim of experiment \cite{Mazur} 
was to show that the DOS in presence of the AA effect exhibits a states-conserving dependence on energy.
It has been found \cite{Mazur} that all states removed from near the Fermi level by the AA effect are found at energies above $U_{co}$ 
in the energy range of $2$ to $3$ times $U_{co}$.
However, the observed states-conserving DOS \cite{Mazur} was not compared with theory, because the relevant theories \cite{Altshuler1,Altshuler3,Altshuler4,LeeRamakrishnan,Imry}
studied the AA effect in the low energy limit.
In this work we study the AA effect beyond the low energy limit theoretically.
We consider the model \cite{Altshuler4,Imry}
in which the electrons interact via the statically screened Coulomb interaction and the modification of the DOS is due to the Fock part of the self-energy.
 We derive the states-conserving DOS which is in qualitative accord with experimental observations at energies above $U_{co}$ 
 and which reproduces the AA theory at low energies.
We show that, besides the direct experimental study of the states-conserving DOS \cite{Mazur}, 
such DOS was present (but not noticed) also in other experiments \cite{Schmitz1,Schmitz2,Escudero}.

In our model \cite{Altshuler4,Imry} electrons in the disordered metal interact via the static finite-ranged potential $V(\vec{r}-\vec{r'})$. If  $V = 0$,
 the electrons interact only with the random potential $V_d(\vec{r})$, produced by disorder.
In such case the electron energies $E_m$ and wave functions $\varphi_m$ obey the Schrodinger equation $H \varphi_m(\vec{r}) = E_m \varphi_m(\vec{r})$, 
where $H = - (\hbar^2/2m) \Delta_{\vec{r}} + V_d(\vec{r})$. 
If we treat the e-e interaction within the first order perturbation theory and consider only the Fock part of the interaction,
 $E_m$ is modified to $\tilde{E}_{m}$ as \cite{Imry}
\begin{equation}\label{Fock correction}
     \tilde{E}_{m}  =  E_{m}+\Sigma_{m}^{x}
\end{equation}
 where $\Sigma_{m}^{x}$ is the Fock first-order self-energy correction:
 \begin{equation}\label{exchange energy}
  \Sigma_{m}^{x} =  - \sum_{n} f_{n}  \int  \frac{d\vec{q}}{(2\pi)^{3}} \ V(q) \mid \langle  \varphi_{m} \mid e^{i\vec{q}\cdot\vec{r}}\mid  \varphi_{n} \rangle\mid ^{2}  .
\end{equation}
Here $V(q)$ is the Fourier transform of $V(\vec{r}-\vec{r'})$, $f_n$ is the Fermi function,
 and $\sum_{n}$ is the sum over $n$ with spin parallel to that of $m$.
Equations \eqref{Fock correction} and \eqref{exchange energy} hold if $\Sigma_{m}^{x} \ll E_m$.

Equations \eqref{Fock correction} and \eqref{exchange energy} describe a specific disordered sample. When averaged over many disordered samples,
they remain unchanged except that $E_m$ and $\Sigma_{m}^{x}$ are the mean values. 
Most important, the disorder-averaged $\mid \langle  \varphi_{m} \mid e^{i\vec{q}\cdot\vec{r}}\mid  \varphi_{n}\rangle\mid ^{2}$
can be calculated explicitly. For a diffusing electron \cite{Altshuler4,Imry,LeeRamakrishnan}
\begin{equation}\label{square matrix element}
 \mid \langle  \varphi_{m} \mid e^{i\vec{q}\cdot\vec{r}}\mid  \varphi_{n}\rangle\mid ^{2}=\frac{1}{\pi \rho(E_{n})\Omega}   \ \frac{\hbar D q^{2}}{(\hbar D q^{2})^{2}+(E_{m}-E_{n})^{2}}
\end{equation}
where $D$ is the diffusion coefficient, $\Omega$ is the volume, and $\rho(E_{n})$ is the DOS for a single spin orientation
[$\rho(E_{n})$ is often replaced by $\rho(E_F)$ which is justified for $E_{n}$ close to $E_F$]. If we average $\tilde{E}_{m}$, $E_{m}$, and $\Sigma_{m}^{x}$
over all states $m$ with energies $E_m = E$, equation \eqref{Fock correction} can be rewritten as
\begin{equation}\label{perturbed mean energy}
    \widetilde{E}(E) = E+ \Sigma^{x}(E) \ ,
\end{equation}
where
\begin{equation}\label{mean selfenergy}
   \Sigma^{x}(E)= - \int_{0}^{E_{F}} dE'\int \frac{d\vec{q}}{8 \pi^{4}} \ V(q) \frac{\hbar D q^{2}}{(\hbar D q^{2})^{2}+(E-E')^{2}}\ .
\end{equation}
In the last equation and in all following calculations we assume zero temperature for simplicity.

Due to averaging over disorder the unperturbed DOS (per spin) reads $\rho_0(E) = {(m/2)}^{3/2} \sqrt{E}/\pi^2 \hbar^3$, as for the free electrons. 
The perturbed DOS
versus $E$, $dn/d\widetilde{E} \equiv \rho(E)$, can be expressed
 \cite{Altshuler4,Imry,LeeRamakrishnan} from equation \eqref{perturbed mean energy} as
\begin{equation}\label{interacting3DdensityatEFselfvesrsusE}
\rho(E) = \rho_{0}(E)\frac{1}{1 + \frac{d \Sigma^{x}(E)}{d E}}   \simeq \rho_{0}(E_F)\frac{1}{1 + \frac{d \Sigma^{x}(E)}{d E}} .
\end{equation}
where the right hand side holds for $\rho_{0}(E) \simeq \rho_{0}(E_F)$.
Note that the perturbed DOS, $\rho(E)$, is expressed as  a function of $E$ rather than of $\widetilde{E}$.
This approximation is valid within the first order perturbation theory \cite{Altshuler4,Imry,LeeRamakrishnan}

It is customary to change the integral $\int_0^{E_F} dE'$ in equation \eqref{mean selfenergy} as $\int_{-\infty}^{E_F} dE'$. 
This \emph{infinite band} approximation is justified for weak interaction. 
Then, substituting $E$ by variable $\varepsilon = E - E_F$, one can rewrite equation \eqref{mean selfenergy} as \cite{LeeRamakrishnan}
\begin{equation}\label{selfenergy relation}
    \Sigma^{x}(\epsilon)=-\int_{\epsilon}^{\infty}d\epsilon' \frac{d \Sigma^{x}(\epsilon')}{d\epsilon'} \ ,
\end{equation}
where \cite{Altshuler4,Imry,LeeRamakrishnan}
\begin{equation}\label{density of states correction}
    \frac{d \Sigma^{x}(\epsilon)}{d\epsilon}= \int \frac{d\vec{q}}{8 \pi^{4}} \  V(q)\  \frac{\hbar D q^{2}}{(\hbar D q^{2})^{2}+\epsilon^{2}}\ .
\end{equation}
The AA effect was studied \cite{Altshuler4,Imry,LeeRamakrishnan} for $V(q)$ so small that $d \Sigma^{x}(\epsilon)/d\epsilon \ll 1$. Then
$\rho(E) \simeq \rho_{0}(E_F)[1 - d \Sigma^{x}(E)/d E]$, or
\begin{equation}\label{interacting3DdensityatEFself}
\rho(\epsilon) \simeq \rho_{0}(0)[1 - d \Sigma^{x}(\epsilon)/d \epsilon] \ .
\end{equation}
In the simplest model \cite{Altshuler4,Imry,LeeRamakrishnan} with static screening
\begin{equation}\label{screenedpotential}
V(q) = \frac{e^2}{\varepsilon_{\infty} \left( q^2  + k_s^2 \right)} \ ,
\end{equation}
where $k_s = \sqrt{e^2 2 \rho_0(E_F)/\varepsilon_{\infty}}$
is the reciprocal screening length (the factor of $2$ is due to the spin degeneracy), and $\varepsilon_{\infty}$ is the high-frequency permittivity of the metal.

From equations \eqref{density of states correction} and \eqref{interacting3DdensityatEFself} one obtains the result of Altshuler and Aronov, 
\cite{Altshuler1,Imry}
\begin{equation}\label{AAdensityofstates3D}
    \rho(E) = \rho(E_F) + \frac{1}{4 \sqrt{2} \pi^2} \frac{ |\epsilon|^{1/2}}{(\hbar D)^{3/2}} ,
\end{equation}
where $\rho(E_F)$ is the DOS at the Fermi level and the second term on the right hand side is the AA interaction correction.
In Refs. \cite{Altshuler1,Imry} the integral in equation \eqref{density of states correction} was calculated assuming $V(q) \simeq V(0)$.
 For $V(q) = V(0)$ the integral
 diverges in the upper limit, therefore, the upper limit was restricted to $q_{max} = \sqrt{\mid\epsilon\mid/ \hbar D}$. 
 Since $V(q) \simeq V(0)$ only for $q \lesssim k_s$,
 the obtained result [equation \eqref{AAdensityofstates3D}] holds only for $\mid\epsilon\mid \lesssim \hbar D k_s^2$. 
 Within this approach the
 term $\rho(E_F)$ remains undetermined, it is usually determined experimentally \cite{Kobayashi,Luna2015}.

In the following text we present an alternative derivation which is not restricted to the low-energy limit. 
At low energies our derivation will reproduce equation \eqref{AAdensityofstates3D} and also determine explicitly the term $\rho(E_F)$. 
However, our major goal is to go beyond the low energy limit 
and to derive the DOS which conserves the states similarly as in the experiment \cite{Mazur}.

In Sect. 2 we show that the DOS given by equations \eqref{interacting3DdensityatEFself}, \eqref{density of states correction}, and \eqref{screenedpotential} does not conserve the states.
In Sect. 3. we identify why this is so and present a heuristic derivation of the states-conserving DOS.
 Comparison with experiment is presented in Sects. 3 and 4. Finally, in Sect. 5 we interpret the AA effect with conservation of states in terms of coupling 
 between the interaction \eqref{screenedpotential} and matrix element \eqref{square matrix element}.

\section{The states conservation problem}

\begin{figure}[t]
\centerline{\includegraphics[clip,width=0.85\columnwidth]{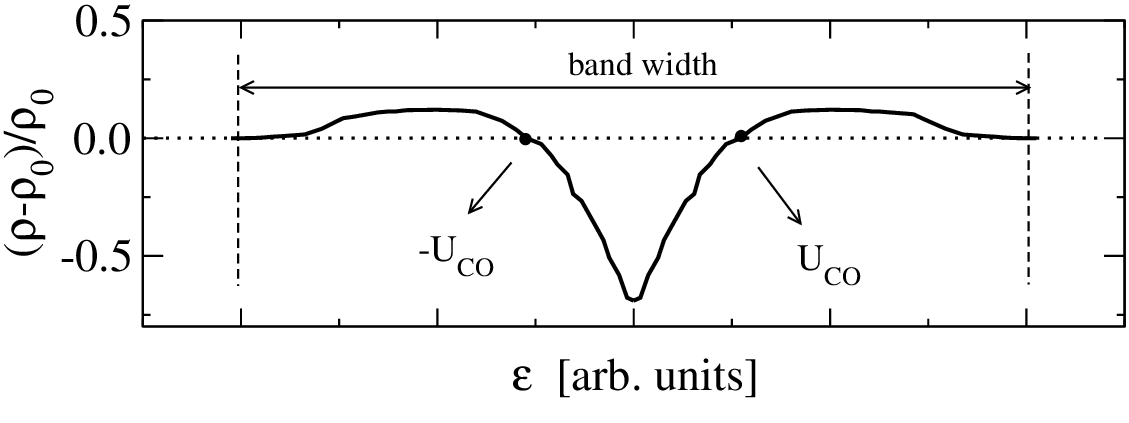}}
\caption{Experimental output (schematic) for DOS in a weakly disordered metal, normalized as $[\rho(\epsilon)-\rho_0(0)]/\rho_0(0)$. 
The correlation energy $U_{co}$, defined \cite{Mazur} by equation $\rho(\epsilon)=\rho_0(0)$, is marked by arrow. 
The states conservation means that $\mid \int_0^{U_{co}} d\epsilon (\rho - \rho_0)/\rho_0 \mid \ =
 \ \int_{U_{co}}^{\infty} d\epsilon (\rho - \rho_0)/\rho_0$, assuming that the conduction band width is much larger than the states conservation region.
}
\label{Fig:1}
\end{figure}

Figure \ref{Fig:1} shows schematically the typical experimental output \cite{Mazur}. At energies below $U_{co}$ the data show the AA singularity 
described by the $|\epsilon|^{1/2}$ law.
All states repelled from the AA singularity are found at energies above $U_{co}$ in the range of about $3 U_{co}$.
This \emph{local} conservation of states should be distinguished from the conservation of states in strongly correlated disordered systems 
where the states repelled by interaction are transferred far away from the
Fermi level \cite{Chen}. In the latter case one cannot use the approximation of the infinitely wide band which on the contrary has no effect 
if conservation of states takes place \emph{locally} near the Fermi level.
Whenever we speak about the conservation of states, we have in mind the \emph{local} conservation of states similar to that in figure \ref{Fig:1}.

In accord with figure \ref{Fig:1} and Ref. \cite{Mazur}, the conservation of states for the AA model reviewed in Sect. 1 reads

\begin{equation}\label{conservationofstates}
\int_0^{\infty} d \epsilon [\rho(\epsilon) - \rho_0(0)]=0 \ .
\end{equation}
Inserting equation \eqref{interacting3DdensityatEFself} into the conservation law \eqref{conservationofstates} we find that
the conservation of states is fulfilled only if
\begin{equation}\label{conservationofstates generalcondition}
    \int_{0}^{\infty}d\epsilon \frac{d \Sigma^{x}(\epsilon)}{d\epsilon} = 0 \ .
\end{equation}
 However, equation \eqref{conservationofstates generalcondition} is not fulfilled  because $d \Sigma^{x}(\epsilon)/d\epsilon$ 
 is positive for any $\epsilon$ [see equation \eqref{density of states correction}].
This means that the model of Sect. 1 does not conserve the states.

Furthermore, integral $\int_{0}^{\infty}d\epsilon \ d \Sigma^{x}(\epsilon)/d\epsilon $ not only fails to fulfill equation \eqref{conservationofstates generalcondition} 
 but even  diverges in the upper limit (see the next section).
  This means that also the self-energy \eqref{selfenergy relation} diverges which is another problem, in addition to the states conservation problem. 
  In principle, the divergence could be eliminated by considering the energy band of finite width, 
  however, the self-energy would then depend on the band width which is also not sound (for weak interaction we expect 
  $\Sigma^x \lesssim e^2/4 \pi \varepsilon_{\infty}k_s^{-1}$
  independently on the band width).    
  We will see that the divergent self-energy is closely related to the conservation of states problem.

To conserve the states and to obtain a finite self-energy, one has to identify the limitations of the model and to modify it properly. 
First, the diffusive approximation \eqref{square matrix element} holds only at small energies. Second, the
interaction \eqref{screenedpotential} and diffusive matrix element \eqref{square matrix element} are mutually independent while in reality 
they should affect each other.

Concerning the first point, the diffusive approximation \eqref{square matrix element} holds for the eigen-states which are correlated in time and space \cite{Altshuler4,Imry,LeeRamakrishnan}. 
Therefore, equation \eqref{square matrix element} is valid only if
$\mid E_{m}-E_{n} \mid \lesssim \hbar/\tau$ and $q \lesssim 1/v_F\tau$, where $\tau$ is the elastic scattering time and $v_F$ is the Fermi velocity.
In addition, the e-e interaction
with spatial range $\sim 1/k_s$ introduces the correlation time $\tau_{co} \sim k_s^{-2}/D$ which is usually longer than $\tau$ and gives 
rise to the correlation energy $U_{co} \sim \hbar D k_s^{2}$.
This restricts the validity of equation \eqref{square matrix element} to even smaller $\mid E_{m}-E_{n} \mid$, say to  $\mid E_{m}-E_{n} \mid \lesssim U_{co}$. 
For $\mid E_{m}-E_{n} \mid  > U_{co}$ the matrix element \eqref{square matrix element} has to be modified by e-e interaction and to depend on $U_{co}$. 
Due to the decaying correlation between $m$ and $n$, it should decay with increase of $\mid E_{m}-E_{n} \mid$ faster than now.

Second, in a realistic self-consistent model the interaction \eqref{screenedpotential} has to be modified to the form which depends on the energy difference
$\mid E_{m}-E_{n}  \mid $ and diffusion coefficient $D$. From figure \ref{Fig:1} it follows that the conservation of states 
(equation \ref{conservationofstates}) can be fulfilled only if
$d\Sigma^{x}(\epsilon)/d\epsilon$ in equation \eqref{conservationofstates generalcondition} changes the sign at energy $ \mid \epsilon \mid  = U_{co}$. 
Indeed, one can see from equation \eqref{density of states correction} that such sign change
can arise only from the sign change of the interaction $V$.

\section{The states-conserving DOS and self-energy}

The question is how to modify equation \eqref{density of states correction}
to be valid also beyond the low energy limit. A direct self-consistent solution of the problem would be difficult. Therefore, as a first approach, we develop a simple heuristic theory.
Inserting for $V(q)$ the equation \eqref{screenedpotential}, using substitutions $a=\sqrt{\mid\epsilon\mid/\hbar D k_{s}^2}$ and $x=q/k_{s}$, and performing a simple algebra, we
rewrite equation \eqref{density of states correction} as
\begin{equation}\label{density formula continued}
    \begin{split}
       \frac{d \Sigma^{x}(\epsilon)}{d\epsilon}  &  = \frac{1}{ \pi \hbar D k_{s}^2} \frac{e^2}{4 \pi \varepsilon_{\infty} k_{s}^{-1}} \  {\left[ 1+ \frac{{\mid\epsilon\mid}^2}{\hbar^2 D^2 k_s^4} \right]}^{-1}\\
                                & \times  \frac{2}{\pi} \int_{0}^{\infty}dx \left( \frac{1}{1+x^{2}}- \frac{1}{1+(\frac{x}{a})^{4}}+ \frac{x^{2}}{1+(\frac{x}{a})^{4}}   \right) \ .
    \end{split}
\end{equation}
The right hand side of the last equation is composed of three terms. We will show that the only modification we need is the omission of the third term.

All three integrals in equation \eqref{density formula continued} can be
calculated analytically. Introducing notations $U_{co} = 2\hbar D k_{s}^2$ and $U_i = e^2/4 \pi \varepsilon_{\infty} k_{s}^{-1}$,
we obtain
\begin{equation}\label{exactderivativeofselfenerysplitted}
\frac{d \Sigma^{x}(\epsilon)}{d \epsilon} = \frac{d \Sigma_A^{x}(\epsilon)}{d \epsilon} + \frac{d \Sigma_B^{x}(\epsilon)}{d \epsilon} ,
\end{equation}
 where
 \begin{equation}\label{exactderivativeofselfenerypartA}
\frac{d \Sigma_A^{x}(\epsilon)}{d \epsilon} = \frac{2}{\pi} \frac{U_i}{U_{co}} {\left[ 1+ \frac{4{\mid\epsilon\mid}^2}{U_{co}^2} \right]}^{-1}  \left[ 1 - \sqrt{\frac{\mid\epsilon\mid}{U_{co}}} \right],
\end{equation}
corresponds to the first two terms in equation \eqref{density formula continued} and
\begin{equation}\label{exactderivativeofselfenerypartB}
\frac{d \Sigma_B^{x}(\epsilon)}{d \epsilon} =  \frac{2}{\pi} \frac{U_i}{U_{co}} {\left[ 1+ \frac{4{\mid\epsilon\mid}^2}{U_{co}^2} \right]}^{-1} \frac{1}{\sqrt{2}}\left(\frac{2\mid\epsilon\mid}{U_{co}} \right)^{3/2}
\end{equation}
corresponds to the third term.

Using equations \eqref{exactderivativeofselfenerypartA} and \eqref{exactderivativeofselfenerypartB} we find that
\begin{equation}\label{checkconservationofstatesA}
\int_0^{\infty} d \epsilon  \frac{d \Sigma_A^{x}(\epsilon)}{d \epsilon} = 0 \ , \ \ \ \  \int_{\epsilon}^{\infty} d \epsilon' \frac{d \Sigma_B^{x}(\epsilon')}{d \epsilon'} = \infty \ ,
\end{equation}
where the second integral diverges because in the upper limit $\frac{d \Sigma_B^{x}(\epsilon)}{d \epsilon} \propto \mid\epsilon\mid^{1/2}$.
We can now discuss the self-energy.

Inserting equation \eqref{exactderivativeofselfenerysplitted} into the equation \eqref{selfenergy relation} we obtain
\begin{equation}\label{exactselfenerysplitted}
 \Sigma^{x}(\epsilon)  = - \int_{\epsilon}^{\infty} d \epsilon' \ \frac{d \Sigma_A^{x}(\epsilon')}{d \epsilon'} - \int_{\epsilon}^{\infty} d \epsilon' \ \frac{d \Sigma_B^{x}(\epsilon')}{d \epsilon'} \  ,
\end{equation}
where $d \Sigma_A^{x}(\epsilon')/d \epsilon'$ and $d \Sigma_B^{x}(\epsilon')/d \epsilon'$ are given by equations \eqref{exactderivativeofselfenerypartA} and \eqref{exactderivativeofselfenerypartB}.
Due to one of equations \eqref{checkconservationofstatesA} the self-energy \eqref{exactselfenerysplitted} diverges which is not a sound result. A physically sound self-energy should be finite and
of the order of $U_i$
(in our weak-interaction case).
This can only be achieved if we omit in equation \eqref{exactselfenerysplitted} the integral $\int_{\epsilon}^{\infty} d \epsilon' \ d \Sigma_B^{x}(\epsilon')/d \epsilon'$, 
or in other words, if we set $d \Sigma_B^{x}(\epsilon)/d \epsilon \equiv 0$. In fact, we will see soon that the choice $d \Sigma_B^{x}(\epsilon)/d \epsilon \equiv 0$ 
is the only way how to ensure both the finite self-energy and conservation of states. Due to this choice
equation \eqref{exactselfenerysplitted} reduces to  $\Sigma^{x}(\epsilon) =  - \int_{\epsilon}^{\infty} d \epsilon' \ d \Sigma_A^{x}(\epsilon')/d \epsilon'$, 
where the remaining integral is finite and tractable analytically. 
We get
\begin{equation}\label{statesconservingperturbedenergy}
\begin{split}
  & \Sigma^{x}(\epsilon)
     =   \frac{\epsilon}{\mid\epsilon\mid} \    \frac{2}{\pi} \ U_{i}
  \left \{ \frac{\pi}{4} + \frac{1}{8}\ln\left(\frac {1+2\sqrt{\frac{\mid\epsilon\mid}{U_{co}}}+2\frac{\mid\epsilon\mid}{U_{co}}}
  {1-2\sqrt{\frac{\mid\epsilon\mid}{U_{co}}}+2\frac{\mid\epsilon\mid}{U_{co}}}\right)   \right.  \\
  & \left. -\frac{1}{4}\arctan\left(1-2\sqrt{\frac{\mid\epsilon\mid}{U_{co}}}\right)
  -\frac{3}{4}\arctan\left(1+2\sqrt{\frac{\mid\epsilon\mid}{U_{co}}}\right)           \right \} .
  \end{split}
\end{equation}
As expected, the self-energy \eqref{statesconservingperturbedenergy} is finite and smaller than $U_i$.
It starts from zero value at $\epsilon = 0$, reaches the peak value
$\pm 0.16 U_i$ at $\epsilon = \pm U_{co}$, and eventually shows logarithmic decay to zero for $\mid\epsilon\mid \gg U_{co}$.

Now we discuss the DOS. From equations \eqref{exactderivativeofselfenerysplitted} and \eqref{interacting3DdensityatEFself}
\begin{equation}\label{interacting3DdensityatEFselfsplit}
\frac{\rho(\epsilon)}{\rho_{0}(0)}  = 1 - \frac{d \Sigma_A^{x}(\epsilon)}{d \epsilon} - \frac{d \Sigma_B^{x}(\epsilon)}{d \epsilon} \ .
\end{equation}
We have already mentioned the conservation of states problem. To conserve the states, equation \eqref{interacting3DdensityatEFselfsplit} has to fulfill equation \eqref{conservationofstates}. 
So it conserves the states only
if
\begin{equation}\label{checkconservationofstates}
\int_0^{\infty} d \epsilon  \frac{d \Sigma_A^{x}(\epsilon)}{d \epsilon} +  \int_0^{\infty} d \epsilon \frac{d \Sigma_B^{x}(\epsilon)}{d \epsilon} = 0 \ ,
\end{equation}
which is not the case due to the second of equations \eqref{checkconservationofstatesA}.
Evidently, the term $d \Sigma_B^{x}(\epsilon)/d \epsilon$ is not sound due to its behavior at large energies, in accord with the fact that equation \eqref{mean selfenergy} is not valid at large energies.
On the other hand, the first of equations \eqref{checkconservationofstatesA} implies that the term $d \Sigma_A^{x}(\epsilon)/d \epsilon$ is sound in the sense that it conserves the states.
Motivated by equations \eqref{checkconservationofstatesA} and \eqref{checkconservationofstates}, we set in equation \eqref{interacting3DdensityatEFselfsplit}
 $d \Sigma_B^{x}(\epsilon)/d \epsilon \equiv 0$ and
we obtain the DOS $\rho(\epsilon)/\rho_0(0) = 1 - d \Sigma_A^{x}(\epsilon)/d \epsilon$
which conserves the states. In final form
\begin{equation}\label{statesconservingDOS}
\frac{\rho(\epsilon)}{\rho_0(0)} = 1 - \frac{2}{\pi} \frac{U_i}{U_{co}} {\left[ 1+ \frac{4{\mid\epsilon\mid}^2}{U_{co}^2} \right]}^{-1}  \left[ 1 - \sqrt{\frac{\mid\epsilon\mid}{U_{co}}} \right] \ .
\end{equation}

\begin{figure}[t]
\centerline{\includegraphics[clip,width=0.8\columnwidth]{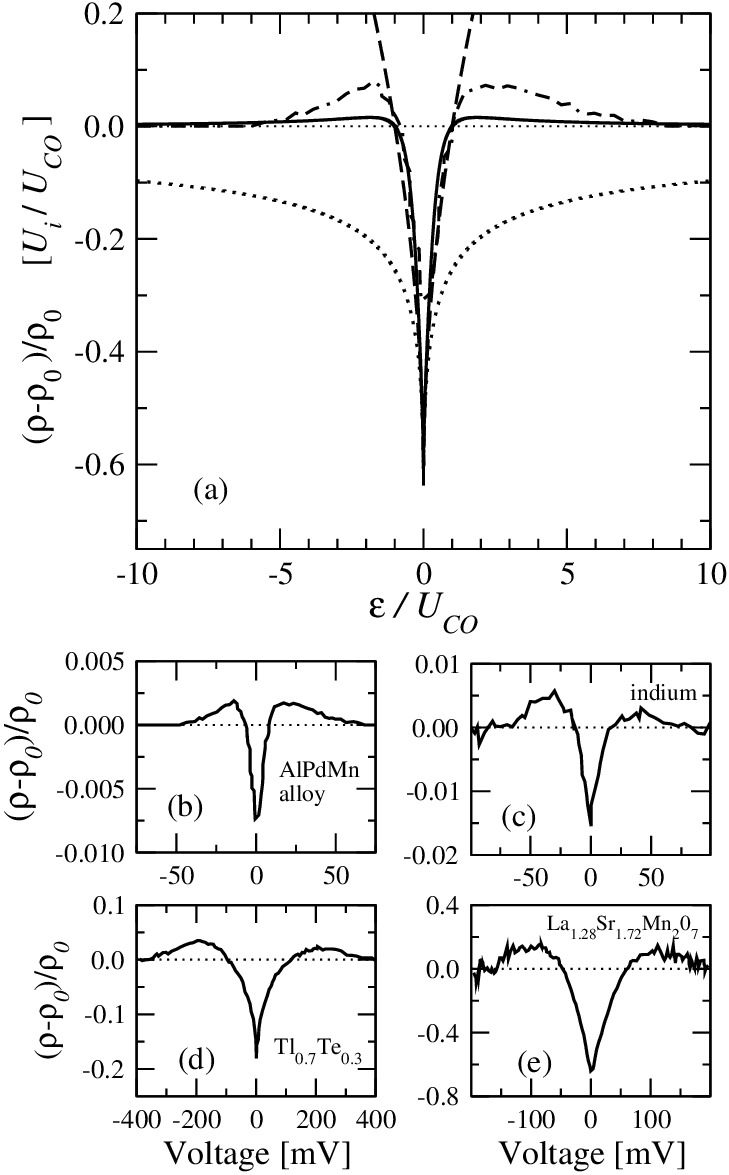}}
\caption{
Panel \emph{a} shows the DOS, $(\rho(\epsilon) - \rho_0(0))/\rho_0$, in units $U_i/U_{co}$ in dependence on the normalized energy $\epsilon/U_{co}$.
 Namely, the full line shows the states-conserving DOS
\eqref{statesconservingDOS}, the dashed line shows the low-energy limit \eqref{lowenergylimitintermsofUiandUco}, and the dotted line shows the result \eqref{relativedensityofstatesour}.
Panels \emph{b},\emph{c},\emph{d}, and \emph{e} show the experimental data for four different disordered metals. 
The data in panels \emph{b}, \emph{c},  and \emph{d} were extracted from the experimental data of Refs. \cite{Escudero}, \cite{Schmitz1}, and \cite{Schmitz2}, 
respectively (see Sect. 4 for details).
The data in panel \emph{e} were taken from Ref. \citep{Mazur} as they are. The data from panel \emph{b} are also shown in units $U_i/U_{co}$ 
in panel \emph{a} (the dotted-dashed line). For these data we roughly estimate $U_{co}= 8$ meV and $U_i =0.19$ meV.
}
\label{Fig:2}
\end{figure}

The DOS expression
\eqref{statesconservingDOS} is plotted in a full line in panel \emph{a} of figure \ref{Fig:2}. The full curve exhibits the AA singularity at energies $\mid\epsilon\mid \lesssim U_{co}$.
 Due to the conservation of states, the states removed
 from the AA singularity are found at energies $\mid\epsilon\mid > U_{co}$.
About one half of states piles up between $\mid\epsilon\mid = U_{co}$ and $\mid\epsilon\mid = 10 U_{co}$, a further one third (not shown) piles up between $\mid\epsilon\mid = 10 U_{co}$ and $\mid\epsilon\mid = 20 U_{co}$.

 For comparison, panels \emph{b}, \emph{c}, \emph{d}, and \emph{e} show the experimental data for various disordered metals,
 the experimental data of panel \emph{b} are shown also in panel \emph{a} (the dotted-dashed line). The origin of all these data is specified in the next section.
 Here we point out the following.

 First, all experimental curves exhibit the AA singularity accompanied by
 pile up of states at energies above $U_{co}$. This is a sign of the \emph{local} conservation of states (Fig. \ref{Fig:1}). 
 Second, the states-conserving DOS  \eqref{statesconservingDOS} (the full curve in panel \emph{a}) captures the main experimental 
 features in the sense that it describes the AA singularity quantitatively (at $T = 0$) and mimics the observed pile up of states qualitatively.
Quantitatively, the width of the pile up of states region in the experiment is much smaller (five to eight times) than in our theory. 
This is likely due to the fact that we ignore the second order interaction effects which can be  
important for large $\mid\epsilon\mid$. Moreover, we
rely on the constant diffusion coefficient $D$ (Sect. 1) while in reality $D$ should decrease with increase of $\mid\epsilon\mid$ 
and even turn to zero due to the Anderson localization.
Finally, experimental data are affected by finite $T$ while we assume $T = 0$.

Equation \eqref{statesconservingDOS} not only conserves the states but in the limit $\mid\epsilon\mid^2 \ll U_{co}^2/4$ gives the formula
\begin{equation}\label{lowenergylimitintermsofUiandUco}
\frac{\rho(\epsilon)}{\rho_0(0)} = 1 - \frac{2}{\pi} \frac{U_i}{U_{co}} + \frac{2}{\pi} \frac{U_i}{U_{co}} \sqrt{\frac{\mid\epsilon\mid}{U_{co}}}  \ ,
\end{equation}
which reproduces the AA result \eqref{AAdensityofstates3D} and in addition expresses $\rho(E_F)$ as
$\rho(E_F) = \rho_{0}(E_F) [ 1 - (2/\pi) (U_i/U_{co}) ]$. In figure \ref{Fig:2}\emph{a} equation \eqref{lowenergylimitintermsofUiandUco} 
is plotted in a dashed line.


\begin{figure}[t]
\centerline{\includegraphics[clip,width=0.7\columnwidth]{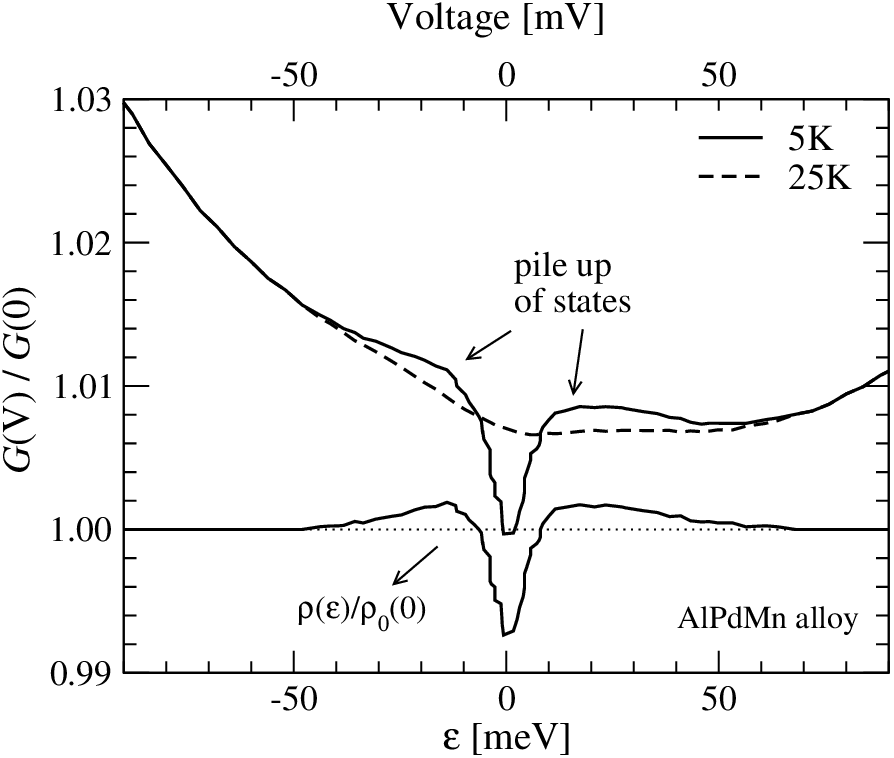}}
\caption{Tunneling spectrum $G(V)$ versus $V$, measured \cite{Escudero} at temperatures $5$ K and $25$ K and normalized by $G(0)$ at $5$ K. 
The data were taken from Fig. 10 of reference \cite{Escudero}, the studied metal was the $AlPdMn$ alloy. 
Also shown is the $\rho(\epsilon)/\rho_0(0)$ dependence at $5$ K, determined (by us) as $\rho(\epsilon)/\rho_0(0) \simeq G(V,5K)/G(V,25K)$.}
\label{Fig:3}
\end{figure}

Finally, we express explicitly equation \eqref{interacting3DdensityatEFselfsplit}:
\begin{equation}\label{relativedensityofstatesour}
\begin{split}
\frac{\rho(\epsilon)}{\rho_0(0)} & = 1 - \frac{2}{\pi} \frac{U_i}{U_{co}} {\left[ 1+ \frac{4{\mid\epsilon\mid}^2}{U_{co}^2} \right]}^{-1}   \\
 & \times  \left[ 1 - \sqrt{\frac{\mid\epsilon\mid}{U_{co}}} +  \frac{1}{\sqrt{2}}\left(\frac{2\mid\epsilon\mid}{U_{co}} \right)^{3/2} \right] \ .
\end{split}
\end{equation}
In figure \ref{Fig:2}\emph{a} the DOS expression \eqref{relativedensityofstatesour} is shown in a dotted line, 
a similar dependence was found in work \cite{Rabatin}.
It shows no pile up of states as it does not conserve them. It also
deviates too early from the AA singularity because it reduces to the low energy limit \eqref{lowenergylimitintermsofUiandUco}
for $\mid\epsilon\mid \ll U_{co}/2$ rather than for $\mid\epsilon\mid^2 \ll U_{co}^2/4$. The former limit is more restrictive
and this restriction is due to the term $d \Sigma_B^{x}(\epsilon)/d \epsilon$ in equation \eqref{relativedensityofstatesour}.
This is another reason for choice $d \Sigma_B^{x}(\epsilon)/d \epsilon \equiv 0$.

\section{Insight into experiment}

The experimental data in figure \ref{Fig:2} originate from the tunneling spectroscopy experiments \cite{Escudero,Schmitz1,Schmitz2,Mazur}. 
In these experiments the electron current ($I$) is driven through the metal-insulator-metal (MIM) tunnel junction composed of 
the clean metal and disordered metal of interest. The output is the differential conductance $G(V) \equiv dI(V)/dV$ 
in dependence on voltage $V$.
 At low temperatures
$G(V)$ can be expressed as $G(V)/G_0(V) \simeq \rho(\epsilon)/\rho_0(0)$, where $\epsilon = -eV$ and $G_0(V)$ 
is the differential conductance that is independent on the interaction \cite{Mazur}.
In reality
$G_0(V)$ is a parabolic function which has to be specified \cite{Mazur} if one studies $\rho(\epsilon)$ at large energies.

\begin{figure}[t]
\centerline{\includegraphics[clip,width=0.97\columnwidth]{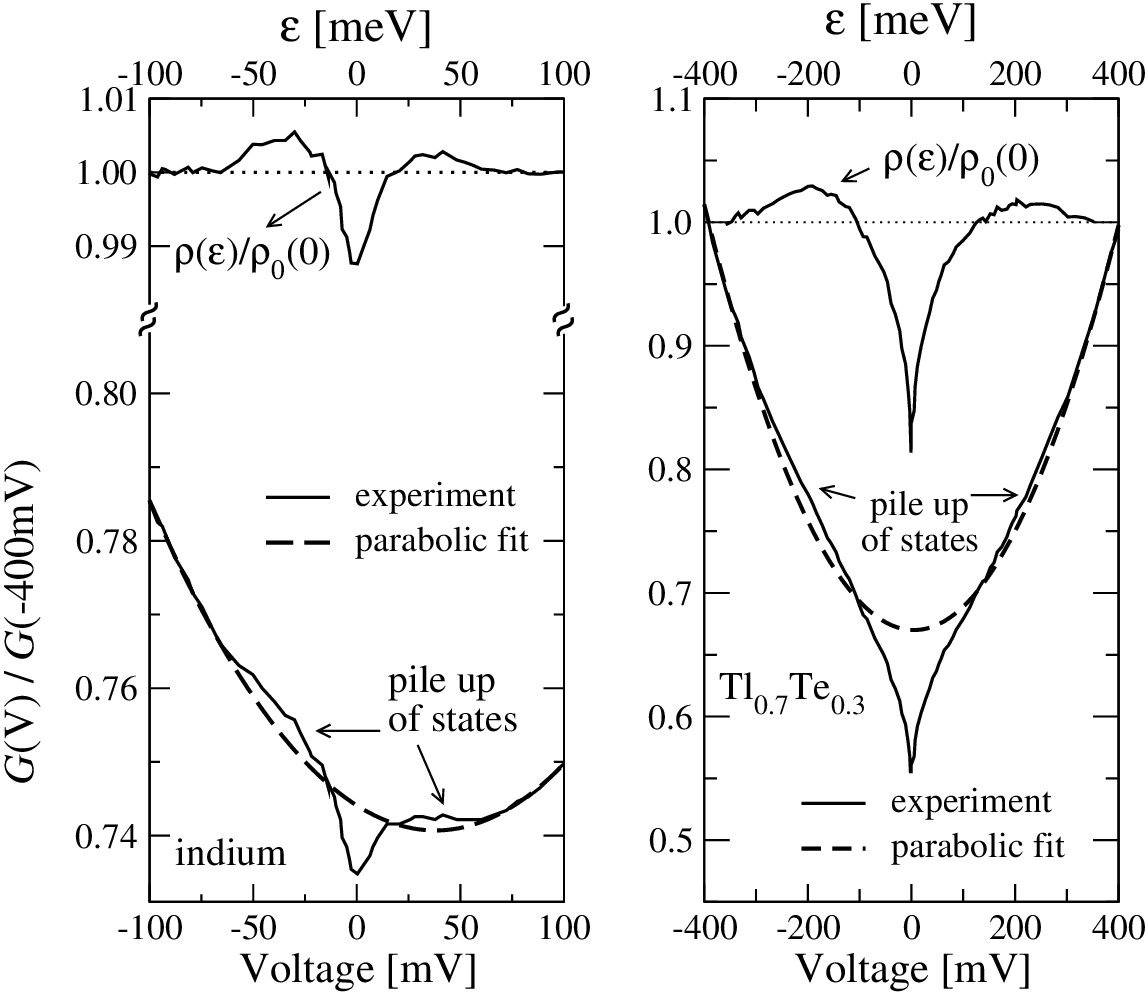}}
\caption{ Tunneling spectrum $G(V)/G(-400 \ mV)$ measured \cite{Schmitz1,Schmitz2} at $4.2$ for indium (left) and $Tl_{0.7}Te_{0.3}$ (right).
 The spectrum on the left originates from figure 1 of work \cite{Schmitz1}, the spectrum on the right from figure 2 of work \cite{Schmitz2}. 
 The spectra exhibit the AA singularity (studied in cited works) and also the pile up of states features (marked by us).
The dashed line is our parabolic fit of the $G(V)$ spectrum away from the pile up features, this fit provides the parabolic $G_0(V)$ background.
 We then use equation $\rho(\epsilon)/\rho_0(0) \simeq G(V)/G_0(V)$ and obtain the plotted DOS curves $\rho(\epsilon)/\rho_0(0)$.
}
\label{Fig:4}
\end{figure}

To our knowledge, experiment \cite{Mazur} is the only experiment which studied the AA effect with intention to observe the pile up of states.
In that work $G(V)$ was measured for the disordered $La_{1.28}Sr_{1.72}Mn_{2}O_{2}$ metal at $4.2$ K and $\rho(\epsilon)/\rho_0$ 
was determined from equation $G(V)/G_0(V) \simeq \rho(\epsilon)/\rho_0$ after specifying $G_0(V)$. 
The resulting $\rho(\epsilon)/\rho_0$ data are those presented in panel \emph{e} of figure \ref{Fig:2}, 
these data represent the first observation \cite{Mazur} of the \emph{local} conservation
of states in the weakly disordered metal. 
Now we show that such \emph{local} conservation of states was present (unnoticed) also in other experiments.

The $\rho(\epsilon)$ dependence for the $AlPdMn$ alloy, shown in panel \emph{b} of figure \ref{Fig:2}, 
has been determined (by us) from the $G(V)$ data measured in work \cite{Escudero}. Figure \ref{Fig:3}  shows
the $G(V)$ curves measured \cite{Escudero} at $5$ K and $25$ K, normalized by $G(0)$ at $5$ K.
 The only difference between these curves is a pronounced AA singularity at $5$ K and almost no AA singularity at $25$ K, otherwise both curves coincide. 
  Thus, the $G(V)$ curve at $25$ K can play the role of $G_0(V)$ and we can
determine $\rho(\epsilon)/\rho_0(0)$ at $5$ K as $\rho(\epsilon)/\rho_0(0) \simeq G(V,5K)/G(V,25K)$.
 The resulting  $\rho(\epsilon)$ is shown in figure \ref{Fig:3}, the same result was used in panel \emph{b }of figure \ref{Fig:2}. 
 The $G(V)$ curve at $5$ K exhibits the peaks which we have marked by arrows.
  These peaks are due to the pile up of states repelled from the AA singularity.

The $\rho(\epsilon)$ dependencies for $In$ and $Tl_{0.7}Te_{03}$, shown in panels \emph{c} and \emph{d} of figure \ref{Fig:2},
 were extracted by us from the $G(V)$ spectra measured in works \cite{Schmitz1,Schmitz2}.
  The spectra are presented in figure \ref{Fig:4} together with our extraction procedure. 
  The $G(V)$ spectra exhibit the AA singularity together with features due to the pile up of states.
   We have extracted from the spectra the parabolic $G_0(V)$ background and we have obtained $\rho(\epsilon)/\rho_0(0)$.

\section{Discussion, alternative view on theory}

Our major result is the states-conserving DOS given by equation \eqref{statesconservingDOS}.
At low energies equation \eqref{statesconservingDOS} reproduces the original AA result \eqref{AAdensityofstates3D}
and determines the term $\rho(E_F)$. 
Most important, at large energies it provides the conservation of states, manifested by pile up of states above the correlation energy $U_{co}$. 
This pile up of states is qualitatively similar to that observed experimentally \cite{Mazur}. 
We have shown that such pile up of states feature was present also in other experiments \cite{Schmitz1,Schmitz2,Escudero}.

Our derivation was heuristic. 
We have found that the AA self-energy consists of a diverging part and of the small part of size $\sim e^2/\varepsilon_{\infty}k_s^{-1}$ (equation \ref{statesconservingperturbedenergy}). 
We have removed the diverging part (on physical grounds) by omitting the term $d \Sigma_B^{x}(\epsilon)/d \epsilon$ 
and the remaining part has produced the states-conserving DOS which we have searched for. 
In this sense the conservation of states is inherent to the AA model.

We want to finish by providing an alternative view on our derivation. 
In the AA model the matrix element \eqref{square matrix element}  and interaction \eqref{screenedpotential} were independent, 
however, in reality they should affect each other. 
We show that just this happens in our model due to the omission of term  $d \Sigma_B^{x}(\epsilon)/d \epsilon$.

We first repeat in a slightly different form our major results. 
We take equation \eqref{density formula continued} and omit the third term on the right hand side
(the third term is $d \Sigma_B^{x}(\epsilon)/d \epsilon$). 
We get 
\begin{equation}\label{density formula rewritten}
    \begin{split}
       \frac{d \Sigma^{x}(\epsilon)}{d\epsilon} & =  \frac{1}{8\pi^{4}}  \ \left[1+\frac{\epsilon^{2}}{\hbar^{2}D^{2}k_{s}^{4}}\right]^{-1}  4 \pi \int_{0}^{\infty} dq  \ q^{2} \\
          & \times  \frac{e^2}{\epsilon_{\infty}(q^{2}+k_{s}^{2})}\ \frac{\hbar D q^{2}}{(\hbar Dq^{2})^{2}+\epsilon^{2}} \left[1-\frac{\epsilon^{2}}{\hbar^{2}D^{2}k_{s}^{4}} \ \frac{k_{s}^{2}}{q^{2}}
              \right]  \\
          & = \frac{2}{\pi} \frac{U_i}{U_{co}} {\left[ 1+ \frac{4{\mid\epsilon\mid}^2}{U_{co}^2} \right]}^{-1}  \left[ 1 - \sqrt{\frac{\mid\epsilon\mid}{U_{co}}} \right] \ ,
    \end{split}
\end{equation}
where the right hand side recalls the final result \eqref{exactderivativeofselfenerypartA}.
If we compare equation \eqref{density formula rewritten} with the original equation \eqref{density of states correction}, we see that
the diffusive approximation \eqref{square matrix element} is modified as
\begin{equation} \label{modification1}
\mid \langle  \varphi_{m} \mid e^{i\vec{q}\cdot\vec{r}}\mid  \varphi_{n} \rangle\mid ^{2} =  \frac{1}{\pi \rho \Omega} \frac{\hbar D q^{2}}{(\hbar Dq^{2})^{2}+\epsilon^{2}}  \ \left[1+\frac{\epsilon^{2}}{\hbar^{2}D^{2}k_{s}^{4}}\right]^{-1}
\end{equation}
and the interaction \eqref{screenedpotential} as
\begin{equation} \label{modification2}
V(q, \epsilon) = \frac{e^2}{\varepsilon_{\infty} (q^{2}+k_{s}^{2})} \left[1-\frac{\epsilon^{2}}{\hbar^{2}D^{2}k_{s}^{4}} \ \frac{k_{s}^{2}}{q^{2}} \right] \ .
\end{equation}
Thus, the matrix element and interaction are no longer independent.
First, equation \eqref{modification1} contains the factor $(1+4\epsilon^{2}/U_{co}^2)^{-1}$
 which has the Lorentzian shape with spread $U_{co} = 2 \hbar D k_s^2$. Due to this factor,
expression \eqref{modification1} decreases with increase of $\mid \epsilon \mid$ faster than the diffusive approximation \eqref{square matrix element}.
This is in accord with expectation (Sect. 2) that the interaction will suppress correlation between the states with $\mid E_{m}-E_{n} \mid  > U_{co}$.
Second, interaction $e^2/\varepsilon_{\infty} (q^{2}+k_{s}^{2})$ is modified by
factor  $\left[1- k_{s}^{2} \epsilon^{2}/\hbar^{2}D^{2}k_{s}^{4}q^{2} \right]$ which depends on energy ($\epsilon \equiv E_{m}-E_{n}$) and which changes the sign of the interaction for $\mid \epsilon \mid > \hbar D k_sq$. 
This is again in accord with argument (Sect. 2) that the conservation of states is only possible if the interaction changes the sign for $\mid \epsilon \mid \sim  U_{co}$.

The energy dependence in equation \eqref{modification2} should not be confused with dynamic screening. If we replace
$V(q) =e^2/\varepsilon_{\infty}(q^2 + k_s^2)$ by the real part of $V(q,\omega) =e^2/[q^2 \varepsilon(q,\omega)]$ where  $\varepsilon(q,\omega)= \varepsilon_{\infty} \left( 1 + \frac{k_s^2}{q^2}\frac{D q^2}{-i\omega+D q^2} \right)$ \cite{LeeRamakrishnan}, the interaction remains positive.

Finally, note that the interaction \eqref{modification2} is closely related to the AA effect [equations \eqref{AAdensityofstates3D} and \eqref{lowenergylimitintermsofUiandUco}].
Indeed, the AA equation \eqref{AAdensityofstates3D} is obtained as it stands if we skip in equation \eqref{density formula rewritten} the factor  $(1+\epsilon^{2}/\hbar^2 D^2 k_s^4)^{-1}$ but keep the factor $\left[1- k_{s}^{2} \epsilon^{2}/\hbar^{2}D^{2}k_{s}^{4}q^{2} \right]$.  Owing to this factor it is not necessary to introduce the upper cutoff used in the original derivation [see the discussion of equation \eqref{AAdensityofstates3D}].

\noindent {\it Acknowledgement} 

This work was supported by the APVV project No. APVV-0560-14 and by grant VEGA 2/0200/14.
We thank Richard Hlubina for useful discussions and for help with equations \eqref{density formula continued}, \eqref{exactderivativeofselfenerypartA}, and \eqref{exactderivativeofselfenerypartB}.

\end{document}